\documentclass[12pt]{article}
\usepackage{epsfig}

\newcommand{\bce}{\begin{center}}
\newcommand{\ece}{\end{center}}
\newcommand{\beq}{\begin{equation}}
\newcommand{\eeq}{\end{equation}}
\newcommand{\bea}{\vspace{0.25cm}\begin{eqnarray}}
\newcommand{\eea}{\end{eqnarray}}

\newcommand{\bsigma}{\mbox{\boldmath $\sigma$}}

\newcommand{\bd}{{\bf d}}
\newcommand{\bfe}{{\bf e}}

\newcommand{\bk}{{\bf k}}

\newcommand{\bn}{{\bf n}}

\newcommand{\bp}{{\bf p}}

\newcommand{\bS}{{\bf S}}

\newcommand{\ba}{\begin{array}}
\newcommand{\ea}{\end{array}}

\newcommand{\etal}{{\sl et al.~}}

\newcommand{\doublespace}{
    \renewcommand{\baselinestretch}{1.6}\large\normalsize}

\def\lsim{\mathrel{\rlap{\lower4pt\hbox{\hskip1pt$\sim$}}
    \raise1pt\hbox{$<$}}}         
\def\gsim{\mathrel{\rlap{\lower4pt\hbox{\hskip1pt$\sim$}}
    \raise1pt\hbox{$>$}}}         

\def\Pom{{\bf I\!P}}

\def\lsim{\mathrel{\rlap{\lower4pt\hbox{\hskip1pt$\sim$}}
    \raise1pt\hbox{$<$}}}         
\def\gsim{\mathrel{\rlap{\lower4pt\hbox{\hskip1pt$\sim$}}
    \raise1pt\hbox{$>$}}}         

\def\Pom{{\bf I\!P}}

  \advance\oddsidemargin  by -1in
  \advance\evensidemargin by -1in
  \advance\topmargin      by -0.5in
\makeindex

\doublespace

\begin{document}


\phantom{.}{\bf \Large \hspace{8.0cm} KFA-IKP(Th)-1996-19 \\
\phantom{.}\hspace{9.7cm}October 1996\vspace{0.4cm}\\ }

\begin{center}
{\bf\sl \huge
Nonvanishing tensor polarization of sea quarks in
polarized deuterons}
\vspace{0.4cm}\\
{\bf
N.N.Nikolaev$^{a),b),c)}$ and W.Sch\"afer$^{b)}$
}
\vspace{.5cm}\\
{\sl
$^{a)}$ITKP der Universit\"at Bonn, Nu{\ss}allee 14-16, D-53115 Bonn\\
$^{b)}$IKP, KFA J\"ulich, D-52425 J\"ulich, Germany\\
$^{c)}$L.D.Landau Institute, Kosygina 2, 1117 334 Moscow, Russia}
\vspace{1.0cm}\\
{\Large
Abstract}\\
\end{center}
We propose two new sources of a substantial tensor polarization of
sea partons in the deuteron: the diffractive nuclear shadowing which
depends on the alignment of nucleons in the polarized deuteron and the
nuclear excess of pions in the deuteron which is sensitive to the
spin state of the deuteron. The corresponding tensor structure function
$b_{2}(x,Q^{2})$ rises towards small $x$ and we predict an about
one per cent tensor asymmetry $A_{2}(x,Q^{2})=b_{2}(x,Q^{2})/
F_{2d}(x,Q^{2})$ which by almost two orders in magnitude exceeds the
effect evaluated earlier in the impulse approximation.
The both mechanisms for tensor structure function break the sum rule
$\int dx b_{1}(x,Q^{2})=0$ suggested by Close and Kumano.
We comment on the impact of tensor polarization on the determination
of the vector spin structure function $g_{1d}(x,Q^{2})$ for the
deuteron.
\bigskip\\

\pagebreak

The wide spread theoretical prejudice supported to a certain extent
by experiment is that at high energies which in deep inelastic
scattering (DIS) is $x\rightarrow 0$  total cross sections cease to
depend on the beam and target polarizations (hereafter $x,Q^{2}$
are the standard DIS variables). For instance, the SLAC E143 data
\cite{E143} suggest, and QCD evolution analyses of polarized DIS
support \cite{QCDevol}, the vanishing vector polarization of sea
quarks in the deuteron. In this paper we show that sea quarks in
the spin 1 deuteron treated as a proton-neutron bound state 
have a {\sl non-vanishing tensor polarization}
which even rises at $x \rightarrow 0$. \footnote{A brief report on
these results was presented at the Workshop on Future Physics at
HERA \cite{HERAworkshop}}

If $\bS$ is spin-1 operator then the spin tensor $T_{ik}={1\over 2}
(S_{i}S_{k}+S_{k}S_{i}-{2\over 3}\bS^{2}\delta_{ik})$ is the T-even
and P-even observable. Consequently, for DIS of {\sl unpolarized}
leptons the photoabsorption cross sections $\sigma_{T}^{(\lambda)},
\sigma_{L}^{(\lambda)}$ and structure functions
$
F_{2}^{(\lambda)}(x,Q^2)={Q^{2}\over 4\pi^{2}\alpha_{em}}
(\sigma_{T}^{(\lambda)}+\sigma_{L}^{(\lambda)})
$
are of the form
$
\sigma^{(\lambda)} = \sigma_{0}+\sigma_{2}T_{ik}n_{i}n_{k}+...
$
and depend on the spin projection $\bS\bn=\lambda$ onto the target spin
quantization axis $\bn$ \cite{Khachatryan,Jaffe,Artru}. Hereafter we
focus on the $\gamma^*$-target collision axis (the $z$-axis) chosen for
$\bn$, a somewhat more involved case of $\bn$ normal to the collision
axis will be considered elsewhere. The tensor spin structure function
$b_{2}$ can be defined as
\bea
b_2(x,Q^2) =\frac{1}{2} [F_{2}^{(+)}(x,Q^2)+F_{2}^{(-)}(x,Q^2)-
2F_{2}^{(0)}(x,Q^2)]
\nonumber \\
=
{1\over 2}\sum_{f}e_{f}^{2}x[q_{f}^{(+)}(x,Q^2)+q_{f}^{(-)}(x,Q^2)-
2q_{f}^{(0)}(x,Q^2)]\,.
\label{eq:1.2}
\eea
By the parity conservation $\sigma^{(+)}=\sigma^{(-)}$ and $F_{2}^{(+)}
=F_{2}^{(-)}, q_{f}^{(+)}=q_{f}^{(-)}$. Similarly,
$\sigma_{T}^{(\lambda)}$ and $\sigma_{L}^{(\lambda)}$ define the
transverse, $b_{T}=2xb_{1}$, and the longitudinal, $b_L=b_{2}-b_{T}$,
tensor structure functions. The precursor of the structure
function $b_{1}$ has been discussed by Pais in 1967 \cite{Pais}.

Because vector polarized deuterons are always tensor polarized,
in DIS of polarized leptons on the 100 \% polarized deuteron
the experimentally measured vector spin asymmetry equals $A_{1d}= g_{1d}/
F_{1d}^{(\pm)}$. Consequently, the determination of the vector
spin structure
function $g_{1d}$ only is possible if $F_{1}^{\pm}$ were known:
\beq
g_{1d}(x,Q^{2})=A_{1d}F_{1}^{(\pm)}(x,Q^{2})\, .
\label{eq:1.3}
\eeq
One must not overlook the fact that hitherto $F_{1}^{(\pm)},F_{1}^{(0)}$ are completely unknown and all the
deuteron data have been analyzed under the unwarranted assumption
that $F_{1d}(x,Q^{2})$ for the unpolarized deuteron can be used instead
of $F_{1}^{(\pm)}(x,Q^{2})$ and/or the proper average
$\langle F_1^{(\lambda )} \rangle$ for the not 100\% polarized target
in (\ref{eq:1.3}). 
In order to justify this
assumption and use the polarized deuteron data for tests of the
Bjorken sum rule one needs the direct measurements, and the benchmark
theoretical evaluations, of tensor structure functions of the deuteron.

In the impulse approximation (IA) of Fig.~1a discussed ever since one
must perform the Doppler-Fermi smearing 
\cite{Jaffe}
\beq
F_{2}^{(\lambda)}(x,Q^2)=\int {dy\over y}f^{(\lambda)}(y)
\left[F_{2p}({x\over y},Q^2)+F_{2n}({x\over y},Q^2)\right]\,
\label{eq:1.4}
\eeq
subject to the sum rule for the smearing function
\beq
\int {dy\over y}f^{(\lambda)}(y)=1\,,
\label{eq:1.5}
\eeq
where $y$ is a fraction of the lightcone momentum of the deuteron carried
by the nucleon (nonrelativistically $y=1+p_{z}/m_{N}$, where $\bp$
is the nucleon's momentum in the deuteron and $m_N$ is the nucleon
mass). For the pure $S$-wave deuteron $f^{(\pm)}(y)=f^{(0)}(y)$ and
$b_{2}(x,Q^{2})=0$, the admixture of the $D$-wave makes $f^{(\pm)}(y)
\neq f^{(0)}(y)$ but they differ markedly only at values of $\bp$ such
that $f^{(\pm )}(y),f^{(0)}(y)\ll 1$ \cite{Jeschonnek}. As a result, in
the IA the tensor polarization of quarks $A_{2}(x,Q^{2})=b_{2}(x,Q^{2})
/F_{2d}(x,Q^{2})$ is at most several units of $10^{-4}$ for $x\lsim
0.5$ \cite{Jaffe,Hoodbhoy,Umnikov}. Because of the sum rule
(\ref{eq:1.5}) the tensor polarization of sea quarks vanishes at
$x\rightarrow 0$ and the widely discussed sum rule
\beq
\int dx b_{1}(x,Q^{2})=0\,
\label{eq:1.6}
\eeq
holds in the IA. Capturing on the prejudice that spin effects always
vanish at high energy, Close and Kumano \cite{Close}
conjectured that tensor polarization of partons always vanishes at
$x\rightarrow 0$ and suggested that the sum rule (\ref{eq:1.6}) is a
model independent property of DIS off spin 1 targets. Notice, that in a
disguised form the sum rule (\ref{eq:1.6}) first appeared in
\cite{EfremovTeryaev}, see also \cite{Mankiewicz}. For the constraints
on tensor structure functions from $\sigma^{(\lambda)}\geq 0$ see
\cite{Dmitrasinovic}, higher twist contributions to tensor structure
functions are discussed in \cite{Jaffe3}.

In this paper, treating the deuteron as a proton-neutron bound state,
 we report the first beyond-the-IA evaluation
of the tensor polarization of sea partons in the deuteron from
Gribov's inelastic shadowing diagrams of Fig.~1b and 1c \cite{Gribov}
when the hadronic state $X$ produced by the beam
photon on one nucleon of the deuteron regenerates the photon on the
second nucleon leading to a nuclear eclipse effect:
$
\sigma_{tot}(\gamma^*d) = \sigma_{tot}(\gamma^*p) +
\sigma_{tot}(\gamma^*n) - \Delta\sigma_{sh}(\gamma^*d)\, .
$
There are two major polarization dependent contributions to the
eclipse effect. The first one is the diffractive nuclear shadowing
(NSH) when the state $X$ is excited by pomeron exchange (Fig.~1b).
It is sensitive to the alignment of  nucleons in the polarized deuteron
(\cite{Jeschonnek} and references therein).
In the second mechanism the state $X$ is excited by pion exchange
(Fig.~1c). We comment that apart from sea quarks both mechanisms make
gluon densities too dependent on the tensor polarization of the target
deuteron (for early discussion on gluon densities for spin axis $\bn$
normal to the collision axis see \cite{Artru,Jaffe2,Sather})




Hereafter we focus on small $x$ and do not consider negligible IA terms.
We evaluate diffractive NSH  using Gribov's theory \cite{Gribov}
extended to DIS in \cite{NZ91}:
\beq
\Delta F_{sh}^{(\lambda)}(x,Q^2) =
\frac{2}{\pi}\frac{Q^2}{4\pi^2\alpha_{em}}
\int\!d^2\bk_\perp\!\int\!dM^2
S_D^{(\lambda)}(4\bk^2)\left.\frac{d\sigma^{D}(\gamma^*\to
X)}{dtdM^2}\right|_{t=-\bk_\perp^2} \, .
\label{eq:2.1}
\eeq
Here M is the mass of the state X in diffractive DIS $\gamma^*N\to XN$,
$\bk=(\bk_{\perp},k_{z})$ is the momentum of the pomeron, the polarization
dependence comes from the deuteron form factor $S_D^{(\lambda)}(4\bk^2)$.
Following the conventional wisdom, we neglected the spin-flip pomeron-nucleon coupling.
It is customary to describe diffractive DIS in terms of the pomeron structure
function $F_{2\Pom}(\beta)$ and the flux of pomerons in the proton
$\phi_{\Pom}(x_{\Pom})/x_{\Pom}$:
\bea
(M^2+Q^2)\frac{d\sigma^{D}_{T}(\gamma^*\to
X)}{dtdM^2}=
\frac{\sigma_{tot}(pp)}{16\pi}\frac{4\pi^2\alpha_{em}}{Q^2}
\nonumber\\
\cdot
\left[\phi_{\Pom}^{val}(x_\Pom)F_{2\Pom}^{val}(\beta,Q^2)e^{-B_{val}|t|}
+\phi_{\Pom}^{sea}(x_{\Pom})
F_{2\Pom}^{sea}(\beta,Q^2)e^{-B_{sea}|t|}\right]\, ,
\label{eq:2.3}
\eea
where $\beta=Q^{2}/(Q^{2}+M^{2})$ has a meaning of the Bjorken variable
for DIS on pomerons and $x_\Pom=x/\beta$ is a fraction of
the nucleon's lightcone momentum taken away by the pomeron, in the target
rest frame $k_z= x_{\Pom}m_N$. We use
the predictions \cite{GNZ94} for structure functions $F_{2\Pom}^{sea},
F_{2\Pom}^{val}$ and flux factors $\phi_{\Pom}^{val}(x_{\Pom}),
\phi_{\Pom}^{sea}(x_{\Pom})$ for the valence and sea of the pomeron
from the color dipole approach to
diffractive DIS \cite{NZ92,NZ94} which is known to agree well with the H1
and ZEUS data \cite{ZEUSF2Pom,H1F2Pom}. In (\ref{eq:2.3}) we suppressed
a small contribution from diffractive excitation of the open
charm \cite{NZ92,GNZcharm}.
The experimental determinations of diffraction slopes are as yet lacking,
following \cite{NZ94} in our numerical estimates we take $B_{sea}=
B_{3\Pom}=
6$\,GeV$^{-2}$ and $B_{val} = 2B_{sea}$, the results for NSH only weakly
depend on $B_{val},B_{sea}$.

With the input (\ref{eq:2.3}) NSH can be cast in the convolution form
\beq
\Delta F_{sh}^{(\lambda)}(x,Q^2)=\int_x^1
\frac{dx_{\Pom}}{x_{\Pom}}\left(
\Delta n_{val}^{(\lambda)}(x_{\Pom})
F_{2\Pom}^{val}(\frac{x}{x_{\Pom}},Q^2)+
\Delta n_{sea}^{(\lambda)}(x_{\Pom})
F_{2\Pom}^{sea}(\frac{x}{x_{\Pom}},Q^2)\right)\, ,
\label{eq:2.4}
\eeq
where $\Delta n^{(\lambda)}_{i}(x_{\Pom})$ is a nuclear modification of
the pomeron flux functions ($i=val,sea$):
\beq
\Delta n_{i}^{(\lambda)}(x_{\Pom})=
\frac{2}{\pi}\frac{\sigma_{tot}(pp)}{16\pi}\phi_{\Pom}^{val}(x_{\Pom})
\int\!d^2\bk_\perp S_D^{(\lambda)}(4\bk^2)
e^{-B_{i}|t|} \, .
\label{eq:2.5}
\eeq
We recall that diffractive DIS and NSH are leading twist effects
\cite{NZ91,NZ92}. Consequently this mechanism gives a manifestly
leading twist $b_{2}(x,Q^{2})$. In the momentum representation
\bea
S_D^{(\lambda)}(4\bk^2)=\int\!\frac{d^3\bp}{(2\pi)^3}
Tr(\Psi^{\dagger}_{\lambda}(\bp+\bk)\Psi_{\lambda}(\bp))
=4\pi\int\!\frac{d^3\bp}{(2\pi)^3} \Bigl\{u_0(p')u_0(p)(\bd^*\bd)
\nonumber\\
+\frac{{u_0(p')w_2(p)}}{\sqrt{2}}\Bigl[3(\bd^*\hat{\bp})(\bd\hat{\bp})
-(\bd^*\bd)\Bigr]+\frac{{u_0(p)w_2(p')}}
{\sqrt{2}}\Bigl[3(\bd^*\hat{\bp}')(\bd\hat{\bp}')-(\bd^*\bd)\Bigr]
\nonumber\\
+\frac{w_2(p)w_2(p')}{2}\Bigl[9(\bd^*\hat{\bp}')
(\bd\hat{\bp})(\hat{\bp}\hat{\bp}')-3(\bd^*\hat{\bp})(\bd\hat{\bp})
-3(\bd^*\hat{\bp}')(\bd\hat{\bp}')+(\bd^*\bd )\Bigr]\Bigr\}\, ,
\label{eq:2.7}
\eea
where \cite{Gross}
$
 \Psi_{\lambda}(\bp) = \sqrt{\pi}\left(  \sqrt{2}u_0(p)
(\bsigma \bd) +
w_2(p)\left(3(\bsigma \hat{\bp})(\bd \hat{\bp})-(\bsigma \bd)
\right)\right)$,
$\bsigma$ are the Pauli spin matrices, $\hat{\bp} = \bp/|\bp|$ and $\bp$
is the proton's momentum in the deuteron, $\bp'=\bp+\bk$, the S-
and D-wave functions are related to the familiar radial
functions as $u_0(p) = \int_0^\infty dr r u(r) j_0(pr)$
and $w_2(p) = \int_0^\infty dr r w(r)j_2(pr)$, $\bd(\lambda=0)=\bfe_z$
and $\bd(\lambda = \pm 1)=\frac{1}{\sqrt{2}}(\mp\bfe_x-i\bfe _y)$ are
the deuteron polarization vectors. The alignment of
the deuteron and nonvanishing contributions to $b_{2}(x,Q^{2})$ come
only from the S-D wave interference terms $\propto u_{0}(p)w_{2}(p')+
u_{0}(p')w_{2}(p)$ and the small D-wave term $\propto w_2(p)w_2(p')$.




The pion exchange dominates $\gamma^{*}p \rightarrow nX$, contributes
substantially to $\gamma^{*}p \rightarrow pX$ at a moderately small
$x_{\Pom}$ \cite{HERApion,TripleRegge} and is an well established
source of the $\bar{u}$-$\bar{d}$ asymmetry in DIS on the nucleon
\cite{Zoller,Harald,NA51pion,SpethThomas}. The diagram of Fig.~1c for
DIS on the deuteron can be interpreted as a contribution from DIS
on nuclear excess pions in the deuteron. Here we focus on pions because
the pion exchange is long ranged and arguably must be the dominant
mesonic effect. The contribution of excess pions to $\Delta F_{sh}$ can
be cast in the convolution form
\beq
\Delta F_{sh,\pi}^{(\lambda)}(x,Q^2) = -\int_x^1
\frac{dz}{z}\Delta n^{(\lambda)}_{\pi}(z)
\left[\frac{2}{3}F_2^{\pi^{\pm}}(\frac{x}{z},Q^2)+
\frac{1}{3}F_2^{\pi^0}(\frac{x}{z},Q^2)\right]\,,
\label{eq:3.1}
\eeq
where $F_{2}^{\pi}$ is the pion structure function as parameterized,
for instance, in \cite{GRV,Sutton}, and the flux function for the
excess pions in the deuteron equals
\beq
\Delta n^{(\lambda)}_{\pi}(z) =-6 \frac{g_{\pi NN}^2}{4\pi} z^2
\int\!\frac{d^2\bk_\perp}{(2\pi)^2} \frac{G_{\pi
NN}^2(z,t)}{(t-\mu_{\pi}^2)^2}\!\int\!\frac{d^3\bp}{(2\pi)^3}
Tr\{\Psi^{\dag}_{(\lambda)}(\bp')(\bsigma\bk)
\Psi_{(\lambda)}(\bp)(\bsigma\bk))\}\,,
\label{eq:3.2}
\eeq
where $\bk=(\bk_{\perp},k_{z})$ is the momentum of the exchanged pion,
$k_{z}=zm_{N}$ (notice a similarity between $z$ and $x_{\Pom}$),
$t=-(\bk^2_{\perp}+z^2m_N^2)/(1-z)$, $g_{\pi NN}$ is the
$\pi NN$ coupling and $G_{\pi NN}(z,t)$ is the $\pi NN$ form factor.
Evidently, $\Delta F^{(\lambda)}_{sh,\pi}$ is of leading twist.
At $z\ll 1$  relevant to the present analysis the Regge model
parameterization $G_{\pi NN}(z,t) = \exp (-(\Lambda_0^{2}+\alpha'_{\pi}
\log (\frac{1}{z}))(|t|+\mu_{\pi}^2))$ is the appropriate one (for instance,
see \cite{Sergeev}). With $\Lambda_{0}^{2}=0.3$\,GeV$^{-2}$ and the
slope of the pion trajectory $\alpha_{\pi}'=0.9$\,GeV$^{-2}$, the flux
of pions in the nucleon evaluated with the above Regge form
factor is for all the practical purposes indistinguishable from
the result of the lightcone approach \cite{Zoller,Harald}. In
terms of the S- and D-wave functions $u_0,w_2$ we find
\bea
Tr
[\Psi^{\dag}_{(\lambda)}(\bp')(\bsigma\bk)\Psi_{(\lambda)}(\bp)(\bsigma\bk)] =
4\pi
\left\{ u_0(p)u_0(p') \left( 2 (\bk\bd)(\bk\bd^*)-\bk^2\right)\right.~~~~~~
\nonumber\\
+ \frac{u_0(p)w_2(p')}{\sqrt{2}}\left[
3(\bd^*\hat{\bp}')\left(2(\hat{\bp}'\bk)(\bk \bd)-\bk^2(\hat{\bp}'\bd)\right)
- \left( 2 (\bk\bd)(\bk\bd^*)-\bk^2 \right) \right]~~~~~~
\nonumber\\
+\frac{u_0(p')w_2(p)}{\sqrt{2}} \left[
3(\bd^*\hat{\bp})\left(2(\hat{\bp}\bk)(\bk \bd)-\bk^2(\hat{\bp}\bd)\right)
- \left( 2 (\bk\bd)(\bk\bd^*)-\bk^2\right )\right]~~~~~~
\nonumber\\
+\frac{w_2(p)w_2(p')}{2} \left[ 9
(\bd\hat{\bp})(\bd^*\hat{\bp}')\right.
\left(2(\bk\hat{\bp})(\bk\hat{\bp}') -
\bk^2(\hat{\bp}\hat{\bp}')\right)-
 3(\bd^*\hat{\bp}')\left(2(\hat{\bp}'\bk)(\bk
\bd)-\bk^2(\hat{\bp}'\bd)\right)
\nonumber\\
- \left. 3(\bd^*\hat{\bp})\left.\left(2(\hat{\bp}\bk)(\bk
\bd)-\bk^2(\hat{\bp}\bd)\right)
+ 2(\bk\bd)(\bk\bd^*)-\bk^2\right] \right\} ~~~~~~~
\label{eq:3.3}
\eea
In the evaluation of this matrix element we differ from the expression given
in \cite{Thomas}. These authors considered deep inelastic scattering on 
unpolarized deuterons. Obviously, in this case the pure S-wave contribution must be
of the form $u_0 (p) u_0 (p') \bk^2~.$ The expression in \cite{Thomas} does not 
have this form.




Breaking of the Callan-Gross relation $b_{L}(x,Q^{2})=0$ is of the
origin similar to that in the proton structure function. For instance,
$b_{L}$ from excess pions is found replacing $F_{2}^{\pi}$ in
(\ref{eq:3.1}) by $F_{L}^{\pi}(x,Q^{2})$. For the evaluation of $b_{L}$
from diffractive NSH one needs to know $d\sigma^{D}_{L}(\gamma^*\to X)
/dtdM^2$. The experimental data on $d\sigma_{L}^{D}$ are as yet lacking,
here we use the perturbative QCD results \cite{GNZlong}, according to which
$R=d\sigma_{L}^{D}/d\sigma_{T}^{D}$ is very
similar to that for $ep$ and/or $e\pi$ scattering \cite{GNZlong}.



All the calculations have been performed for the realistic Bonn
\cite{Bonn} and Paris \cite{Paris} wave functions which are
representative of the weak and strong D-wave admixture in the deuteron.
Because of the deuteron form factor in the integrand of (\ref{eq:2.1})
and (\ref{eq:3.2})  $\Delta n^{(\lambda)}_{\pi}(z), \Delta
n^{(\lambda)}_{val}(x_{\Pom})$ and $\Delta n^{(\lambda)}_{sea}(x_{\Pom})$
are nonvanishing only at
\beq
x_{\Pom},z \lsim \langle z\rangle \sim {1 \over R_{d}m_N}\sim 0.2 \, ,
\label{eq:5.1}
\eeq
where $R_{d}$ is the deuteron radius. To have a quick impression on the
effect of excess pions notice that at $x \lsim \langle z \rangle$ the
convolution (\ref{eq:3.1}) can be approximated by
\begin{equation}
\Delta F_{sh,\pi}^{(\lambda)}(x,Q^2) =
-\left(\frac{2}{3}F_2^{\pi^{\pm}}(\frac{x}{\langle z\rangle},Q^2)+
\frac{1}{3}F_2^{\pi^0}(\frac{x}{\langle z\rangle },Q^2)\right)
\langle \Delta n^{(\lambda)}_{\pi}\rangle\, .
\label{eq:5.2}
\end{equation}
The numbers of excess pions $\langle \Delta n^{(\lambda)}_{\pi}\rangle
=\int_0^1 \frac{dz}{z}\Delta n^{(\lambda)}_{\pi}(z)$ are cited in
Table 1. The pure S- and (negligibly small) D-wave terms give
$\langle \Delta n^{(\lambda)}\rangle > 0$, i.e., the antishadowing
effect in $\Delta F_{sh}$. The tensor polarization of sea quarks is
proportional to $\langle \Delta n_{\pi}^{(\pm)}\rangle - 
\langle \Delta n_{\pi}^{(0)}\rangle =1.16 \cdot 10^{-2}$, whereas NSH for
unpolarized deuterons is proportional to ${1\over 3}\left(\langle 2
\Delta n_{\pi}^{(\pm)}\rangle +\langle \Delta n_{\pi}^{(0)}\rangle\right)
=0.42 \cdot 10^{-2}$ for the Bonn wave function. Very similar results are found 
for the Paris wave function: $\langle \Delta n_{\pi}^{(\pm)}\rangle - 
\langle \Delta n_{\pi}^{(0)}\rangle =1.22 \cdot 10^{-2}$ and 
$\frac{1}{3}\left( 2 \Delta n_{\pi}^{(\pm)}\rangle +\langle \Delta n_{\pi}^{(0)}\rangle\right)
=0.34 \cdot 10^{-2}$ and in Fig. 2 and Fig. 3 we show the results only 
for the Bonn wave function. For a comparison,
for the number of pions in a nucleon one finds $\langle n_{\pi}\rangle_N
\sim 0.2$ \cite{Zoller,Harald}. In Fig.~2 and Fig.~3 we show the contribution
from excess pions to $\Delta F_{sh}(x,Q^{2})$ and $b_{2}(x,Q^{2})$.
All numerical results are for $Q^{2}=10$\,GeV$^{2}$ and for the
GRV pion structure function \cite{GRV}, we show them for the
experimentally accessible range of $x$.




The useful scale for NSH corrections at small $x$ is set by the deuteron
spin averaged ratios $\Delta r_{\Pom}^{sea}=
\Delta n_{sea}(x_{\Pom})/
\phi_{\Pom}^{sea}(x_{\Pom})$ and $\Delta r_{\Pom}^{val}=
\Delta n_{val}(x_{\Pom})/\phi_{\Pom}^{val}(x_{\Pom})$, where $\Delta n_i (x_\Pom ) 
= \frac{1}{3} \sum_\lambda \Delta n_i^{(\lambda )}(x_\Pom ),~ i =val,sea~.$
We find $\Delta r_{\Pom}^{sea}=0.1, \Delta r_{\Pom}^{val}=0.08$ for both the Bonn and Paris wave functions. Notice that $\Delta r_{\Pom}^{val}\approx \Delta r_{\Pom}^{sea}$ because of $B_{el},B_{3\Pom}\ll R_{d}^{2}$, which is natural. Our estimate of 
NSH for unpolarized deuterons is shown in Fig.~2. Because the pion
effect is small, our results for NSH are close to those of
Ref.~\cite{BaroneDeuteron} in which only the pure diffractive NSH
has been evaluated in the related model for diffractive DIS.

The results for $b_{2}(x,Q^{2})$ for the deuteron are shown in Fig.~3.
To a crude approximation, for the diffractive NSH mechanism
$b_{2}^{sh}(x,Q^{2}) \sim {1\over 3}F_{sh}(x,Q^{2})$. The excess
pions play a much larger r\^ole in $b_{2}(x,Q^{2})$ than in nuclear
shadowing and dominate $b_{2}(x,Q^{2})$ at $x\gsim 0.01$. At smaller
$x$ the diffractive NSH mechanism takes over. In Fig.~2 and Fig.~3
we also show our estimate for $b_{L}(x,Q^{2})$. Fig.~3b shows that the
tensor polarization of sea quarks in the deuteron $A_{2}(x,Q^{2})$
is about 1 per cent effect and rises towards small $x$.



{\bf Summary and conclusions}.

We have identified and evaluated the two new sources of a substantial
tensor polarization of the sea distribution in deuterons. Our analysis
provides a benchmark calculation of the tensor structure function
of the deuteron at small $x$, very similar results must follow
from other models provided that they give an equally good description
of the experimental data on diffractive DIS. The results for $b_2 (x, Q^2 )$
 are stable against the admissible variations of the
deuteron wave function. The both diffractive nuclear
shadowing and pion excess mechanisms are of essentially nonperturbative
origin but give the leading twist tensor structure function
$b_{2}(x,Q^{2})$. To this end, the pion contribution is obviously
GLDAP evolving, the sea structure function of the pomeron also obeys the
GLDAP evolution \cite{NZ94}, the detailed form of scaling violations
in the valence structure function of the pomeron at $\beta
\rightarrow 1$ remains an open issue \cite{GNZcharm,NZRoma} but
the contribution from this region of $\beta$ to $b_{2}(x,Q^{2})$
is marginal. Consequently, QCD evolution of $b_{2}(x,Q^{2})$
must not be any different from that of $F_{2}(x,Q^{2})$, see
also the related discussion in \cite{Artru,Sather}.

Our finding of the rising tensor polarization of sea quarks in the
deuteron invalidates the assumptions invoked in the derivation of
the Close-Kumano sum rule (\ref{eq:1.6}) and we conclude that the
sum rule (\ref{eq:1.6}) is faulty.

The present uncertainties in the wave function of the deuteron,
structure function of the pion and the diffractive structure
function of the proton do not affect our principal conclusion
of a substantial, $\sim 1$\%, tensor polarization of sea quarks
in the deuteron, which is within the reach of future experiments
with polarized deuterons. Regarding the measurements of $g_{1d}$
in polarized $\vec{e}\vec{d},\vec{\mu}\vec{d}$ scattering, our results
justify the substitution of the unpolarized structure function of
the deuteron for the more correct but still unknown $F_{2}^{(\pm)}$
in (\ref{eq:1.3}) at the present accuracy of measurements of the
spin asymmetry $A_{1d}$, but corrections for the tensor polarization
effects will eventually become important for the quantitative
interpretation of high precision
DIS off polarized deuterons in terms of the
neutron spin structure function $g_{1n}$ and for high accuracy
tests of the Bjorken sum rule.

Finally, eqs.~(\ref{eq:2.4}) and (\ref{eq:3.3}) are fully applicable
to calculation of the tensor polarization of gluon densities in
the deuteron polarized along the $\gamma^{*}d$ collision axis in
terms of gluon structure functions of the pomeron and pion. The
results for the gluon structure functions will be presented
elsewhere.
\bigskip\\
\noindent
{\bf Acknowledgments:} We are grateful to V.N.Gribov, J.Speth and
B.G.Zakharov for helpful discussions. NNN thanks Prof. U.Mei{\ss}ner
and D.Sch\"utte for the hospitality at the Institut f. Theoretische
Kernphysik of the University of Bonn. The work of NNN is supported
by the DFG grant ME864/13-1.

\newpage

\begin{figure} [H]
\begin{center}
\epsfig{file=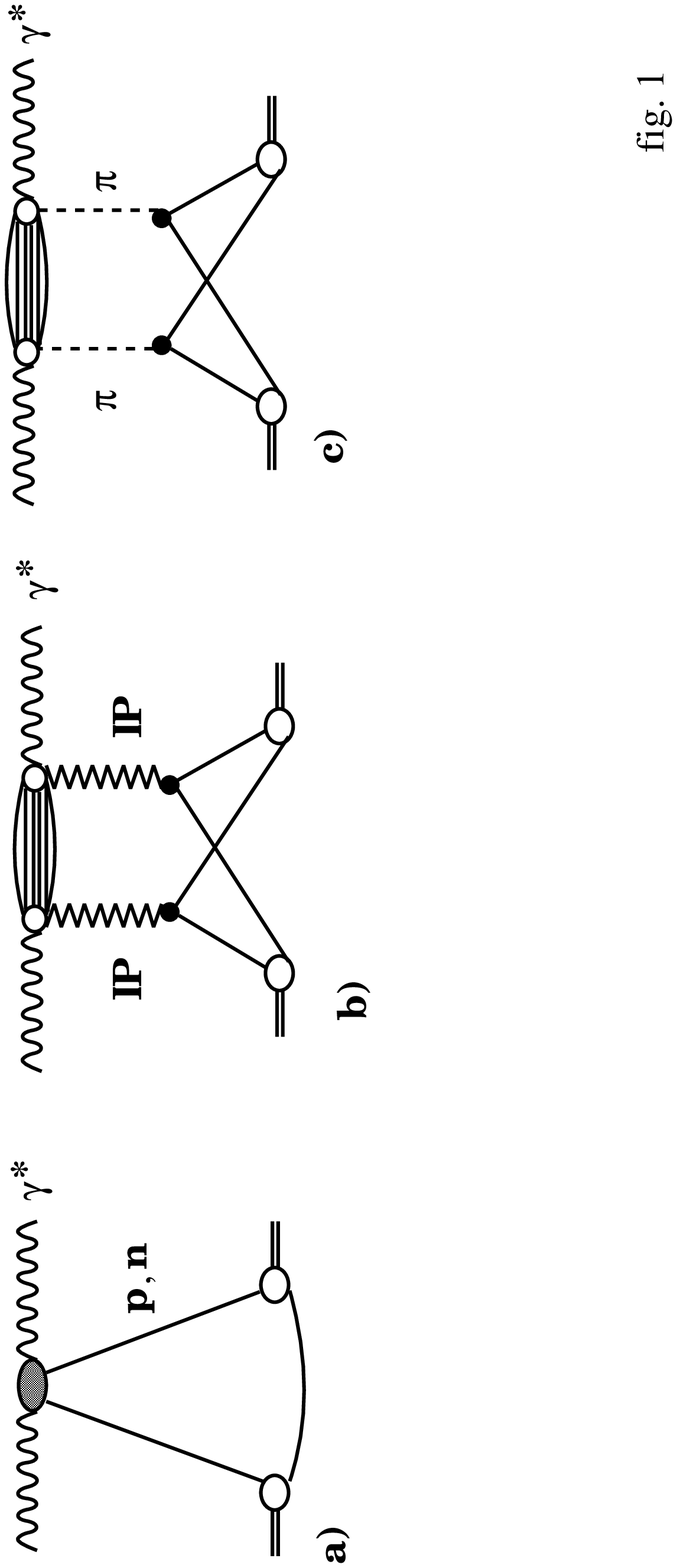, height = 18.0cm}
\end {center}
\end{figure}

\begin{figure} [H]
\begin{center}
\epsfig{file=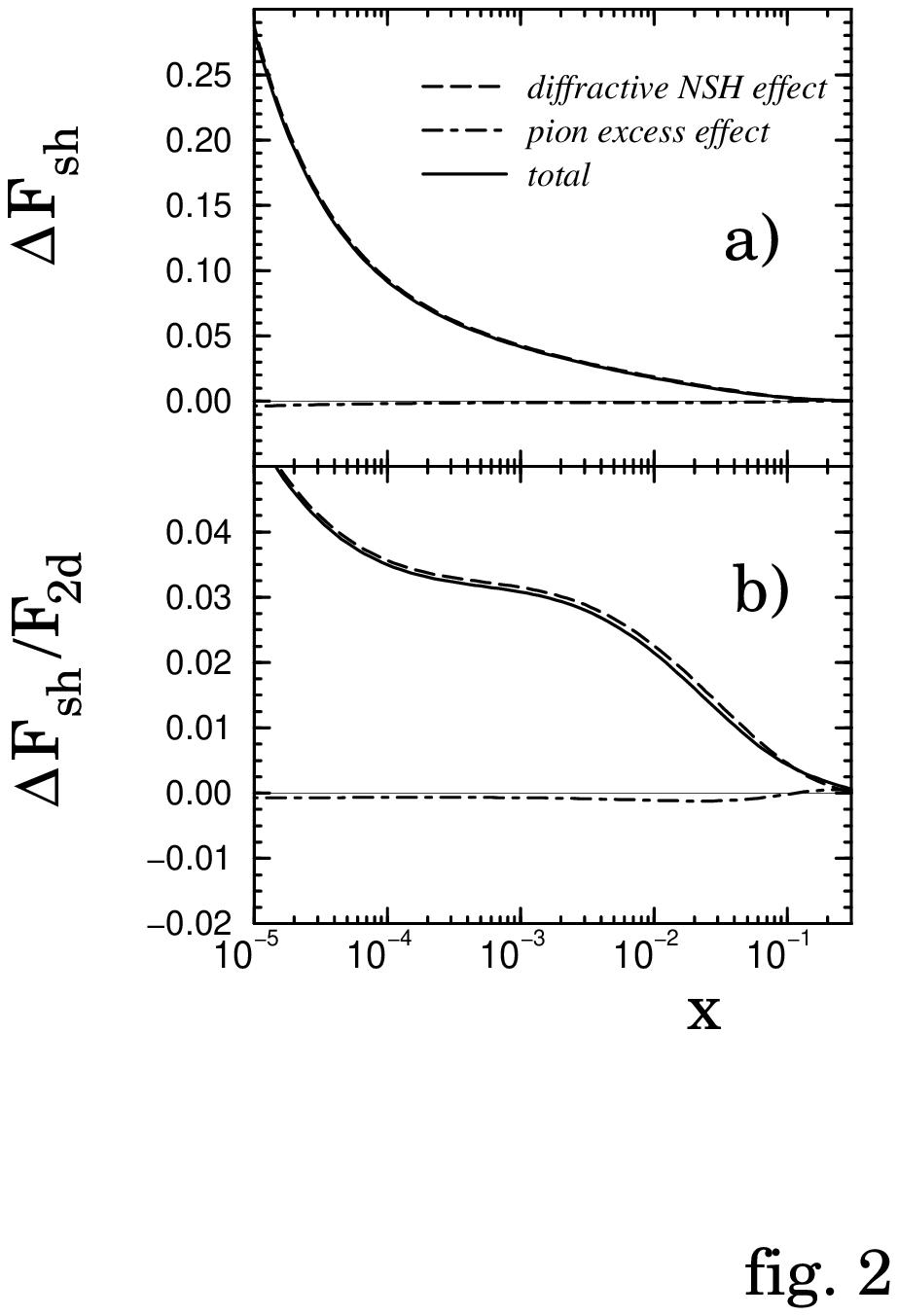, height = 16.0cm}
\end {center}
\end{figure}
\begin{figure} [H]
\begin{center}
\epsfig{file=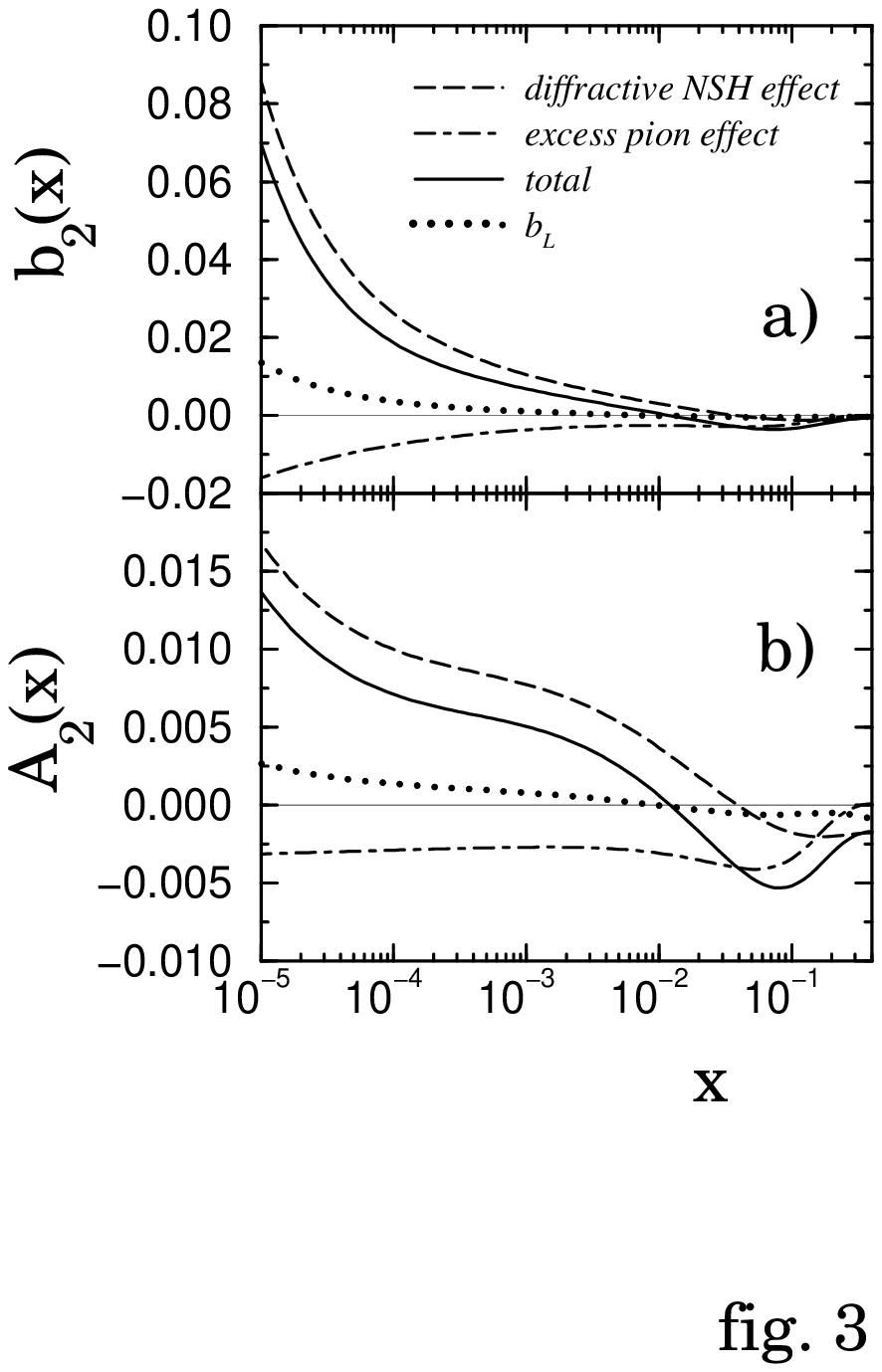, height = 16.0cm}
\end {center}
\end{figure}

\newpage

{\bf{figure captions}}
\begin{itemize}
\item figure 1: {\it Impulse approximation (a) and Gribov's inelastic shadowing (b,c) diagrams for DIS on the deuteron.} 
\item figure 2: {\it (a) Nuclear shadowing correction in the deuteron $\Delta F_{sh} (x,Q^2 ) $ for $Q^2 = 10 GeV^2 $ and its diffractive NSH and pion excess components for the Bonn wave function of the deuteron.
(b) The fractional shadowing correction to the deuteron structure function $F_{2d}(x,Q^2 )$ for $Q^2 = 10 GeV^2~.$}
\item figure 3: {\it(a) The tensor structure function of the deuteron $b_2 (x,Q^2 ) $ at $Q^2 = 10 GeV^2 $ and its diffractive NSH and pion excess components for the Bonn wave function of the deuteron.
(b) The tensor polarization $A_2 (x,Q^2 ) $ of quarks and antiquarks in the deuteron at $Q^2 = 10 GeV^2~.$ Also shown (by the dotted line) is the longitudinal tensor structure function $b_L (x,Q^2 )$ (in (a)), and the corresponding asymmetry (in (b)).}
\end{itemize}

\newpage

\begin{table}
\begin{center}
\begin{tabular}{|l|l|l||l|l|}
\hline
\multicolumn{5}{|c|}{       $\langle \Delta n_\pi^{(\lambda )} \rangle \cdot 10^2$}\\
\hline
\multicolumn{5}{|c|}{       Bonn \hspace{2cm}  Paris}\\
\hline
          & $\lambda =0$  & $\lambda =\pm$ &  $\lambda =0$  & $\lambda =\pm$
  \\
\hline
s-wave    &   0.146       & 0.999         &   0.133       &  0.929
          \\
\hline
s/d-int.  &   -0.537       & -0.268         &   -0.676        &  -0.331
          \\
\hline
d-wave    &   0.035       &  0.085         &  0.070        &  0.153
          \\
\hline \hline
total     &   -0.356       &  0.807         &   -0.466        &  0.750
          \\
\hline
\end{tabular}
\end{center}
\caption{\it The S-D wave decomposition of the mean multiplicities of excess pions in the deuteron for different polarizations $\lambda$ along the $\gamma^* d$--collision axis. The numerical results were obtained for the Bonn and Paris wave functions.}
\end{table}

\end{document}